\documentclass[%
 reprint,
 amsmath,amssymb,
 aps,
]{revtex4-2}

\usepackage{graphicx}
\usepackage{dcolumn}
\usepackage{bm}

\usepackage{amsmath}
\usepackage{ amssymb }
\usepackage{braket}

\begin{document}

\preprint{hi}

\title{Negative Infrared Divergent Corrections to the Global String Tension\\}
\author{Daniel Davies}
\altaffiliation{Santa Cruz Institute for Particle Physics}
\email{dadavies@ucsc.edu}
\affiliation{\small{Department of Physics, University of California, Santa Cruz, Santa Cruz CA, 95064}}

\date{\today}

\begin{abstract}
We show that at large distances from an infinitely long global U(1) string, low energy Goldstone Bosons contribute a spatially dependent term to the energy density which is non-integrable. Specifically, this term diverges with the same form as the classical field theory contribution to the string tension, except with negative sign. This suggests that a careful analysis of the infrared degrees of freedom around the global string ought to yield an effective theory of the solitary string whose tension is finite. There are implications for axion cosmology.
\end{abstract}

\maketitle


\section{The Global String Model}

The model in question has two interacting scalar fields, $\phi_1$ and $\phi_2$, with a potential that is globally $U(1)$ invariant in field space. In complex field notation, with $\Phi = (\phi_1+i\phi_2)/\sqrt{2}$, the tree-level action on 3+1 dimensional Minkowski spacetime with $(+,-,-,-)$ metric signature is

\begin{align}
    S[\Phi]=\int d^4x\left( |\partial\Phi|^2 -V(\Phi^\dagger \Phi)\right)
\end{align}

 We imagine the model is tuned so that its infrared phase admits a spontaneously broken $U(1)$ symmetry, where $\braket{\Phi^\dagger\Phi}=v^2\neq 0$. In other words, the potential may be given by
 
 \begin{align}
     V(\Phi^\dagger\Phi) = \frac{\lambda}{4}(\Phi^\dagger\Phi-v^2)^2
 \end{align}
 
 \noindent In this phase, there are inhomogeneous field configurations with non-trivial topology\cite{1}: they fall into classes defined by elements of the homotopy group $\pi_1(U(1))\cong \pi_1(S_1)=\mathbb{Z}$. The field configurations are known as global strings, and the concern of the present work is the solitary string solution. Assuming the string is straight, infinitely long, and oriented along the $z$-axis, the field configuration is (in cylindrical coordinates)

\begin{align}
    \Phi_n = v f(r) e^{in\theta}
\end{align}

\noindent where $n$ is an integer called the winding number. For large radius we have 

\begin{align}
f(r) \sim 1 - \frac{n^2}{2\lambda v^2 r^2} + \mathcal{O}\left(\frac{1}{r^4}\right)
\end{align}

\noindent while at the origin $f(0)=0$. This solution is easily derived from the equations of motion. The energy density of this profile has large r behavior

\begin{align}
    \mathcal{E} \sim \frac{n^2v^2}{r^2}
\end{align}

\noindent due to the $\theta$-derivative, which is non-integrable on $\mathbb{R}^2$. Therefore the tension of the straight solitary global string is apparently infinite, diverging like $n^2v^2\ln(r_*)$ for some large spatial (infrared) cutoff $r_*$.\\

At the classical level, this means there are no solitary dynamical strings in non-compact spaces. Consequentially, in semi-classical analysis of the global string dynamics, there are always a number of strings such that the total winding number adds to zero, or that the strings are compact; this ensures that the total energy is finite. The strings are assumed to be confined, as the inter-string potential roughly scales logarithmically with the string separation. Any attempt to construct an effective action for such a string network must take careful consideration of the low energy degrees of freedom. This is challenging due to the apparent confining nature of the strings, especially after quantization, because of the technical difficulties surrounding strongly correlated operators at large distances in 4 dimensions. The difficulties with constructing any such perturbative effective theory may be avoided if the classically divergent tension does not survive quantization in the first place. This would imply the inter-string potential scales inversely with separation, which can be easily constructed by weakly coupling the string coordinates to a massless field.\\

\noindent Below the symmetry breaking scale of the model, the degrees of freedom are the background field configuration and the massless Goldstone Bosons of the broken $U(1)$ symmetry, as well as the coordinates of the string. Far below this mass scale, the Goldstones decouple from the massive excitations of the modulus field $|\Phi|$, but if there is a slowly varying background field as in the case with the string, we necessarily must consider the coupling between this background and the massless degrees of freedom. \\

The Goldstone Bosons, which we denote as the field $\alpha$, can be parametrized (ignoring the massive excitations):

\begin{align}
    \Phi = \Phi_n \exp\left(\frac{i\alpha}{v}\right)
\end{align}

\noindent in which case the action for the light sector is
\begin{align}
    S[\alpha] = \int d^4x\:\alpha\left(-v^{-2}\partial_\mu(\Phi^\dagger_n\Phi_n \partial^\mu)\right)\alpha
\end{align}

\noindent where we have made clear the elliptic operator $\Delta = -\partial_\mu(f^2(r) \partial^\mu)$ that will be thoroughly considered in the next section, when we compute the spectrum and therefore the Casimir energy density induced by the vacuum modes of the massless sector at large distances.\\

\section{The Casimir Effect Around a Global String}

Now we will show how the direct calculation of the Casimir energy via summation of the spectrum of the operator $\Delta$ leads to a similar divergence as equation (5). We begin with the Euclideanized operator $\Delta$ (defined by the wick-rotation $\tau = it$), ordered in a perturbative way with powers of $1/r^2$. In other words, we consider the string profile as a perturbation to the dynamics of the Goldstones at very large distances. Using the solution (4) for the string profile we find

\begin{align}
    \Delta = \left(-\nabla^2_{(4)}\right) -\frac{n^2}{\lambda v^2 r^2}\left(-\nabla^2_{(4)} + \frac{2}{r}\partial_r\right) + \mathcal{O}\left(\frac{1}{r^4}\partial_r^2\right)
\end{align}

\noindent where $-\nabla^2_{(4)}$ is the 4-dimensional Laplacian in cylindrical coordinates:

\begin{align}
    -\nabla^2_{(4)} = -\frac{\partial^2}{\partial \tau^2}-\frac{\partial^2}{\partial z^2}-\frac{\partial^2}{\partial r^2} - \frac{1}{r}\frac{\partial}{\partial r}-\frac{1}{r^2}\frac{\partial^2}{\partial \theta^2}
\end{align}

\noindent The eigenfunctions and eienvalues of the leading (unperturbed) operator are

\begin{align}
    \Psi_\xi &= e^{i\omega \tau+iqz+i\ell \theta}J_\ell(kr)\\
     \xi &= k^2+q^2+\omega^2
\end{align}

\noindent These do not constitute a diagonal basis for $\Delta$ for any finite $r$, but they do form a complete orthogonal set of functions with which one should be able to construct solutions at all $r$. Via perturbation theory we find the first order correction to the eigenvalues to be

\begin{align}
    \delta \xi = -\int d^4x \frac{n^2}{\lambda v^2 r^2}J_\ell(kr)\left(-\nabla^2_{(4)} + \frac{2}{r}\partial_r\right)J_\ell(kr)
\end{align}

\noindent where it is understood that the integrand is a valid approximation only for $r$ much larger than the soliton core radius, which is of order $(\lambda v^2)^{-1/2}$. The integral is technically divergent along the $z$ and $\tau$ directions, but this can be ameliorated in the usual way of constructing a wave packet that is localized along those directions. Despite our ignorance about the global nature of the profile, we press onward. The Casimir energy can be derived from the path integral:

\begin{align}
    \int\mathcal{D}\alpha\exp\left(-\int d^4x\; \alpha \Delta \alpha\right) = \det(\Delta)^{-1/2}\\
    = \exp\left(-\frac{1}{2}\text{Tr}\ln\Delta\right)\equiv \exp(-W)
\end{align}

\noindent (ignoring overall multiplicative constants to the partition function which are of no regard) where in our perturbative scheme, the effective action $W$ may be evaluated quite easily:
\begin{widetext}
\begin{align}
    W &= \frac{1}{2}\text{Tr}\ln(\xi+\delta\xi) \sim W_0 + \sum \frac{\delta\xi}{2\xi}\\
    &\sim W_0 -\sum_{\ell=-\infty}^\infty \int d\vec{k} \int d^4x \frac{n^2}{2\lambda v^2 r^2(k^2+q^2+\omega^2)}\Psi^\dagger_\xi(\vec{x})\left(-\nabla^2_{(3)} + \frac{2}{r}\partial_r\right)\Psi_\xi(\vec{x})\\
    &\sim W_0 -\sum_{\ell=-\infty}^\infty\int d\vec{k} \int d^4x \frac{n^2}{2\lambda v^2 r^2(k^2+q^2+\omega^2)}J_\ell(kr)\left(k^2+q^2+\omega^2 + \frac{2}{r}\partial_r\right)J_\ell(kr)
\end{align}
\end{widetext}
\noindent We have used the shorthand notation $d\vec{k}$ to indicate integration over the momentum space spanned by $k,q,\omega$. It is an integral of dimension $\text{mass}^4$. The final component of the integrand in equation (17), where the radial derivative has been left unevaluated, vanishes after summation over $\ell$. The result is compactly expressed:

\begin{align}
    W \sim W_0 -\int d\vec{k} \int d^4x \frac{n^2}{2\lambda v^2 r^2}\sum_{\ell=-\infty}^\infty J^2_\ell(kr)
\end{align}

\noindent Finally, we use the property of Bessel functions of the first kind:

\begin{align}
    \sum_{\ell=-\infty}^\infty J^2_\ell(kr) = 1
\end{align}

\noindent and see that the integrand is completely independent of $k,q$ or $\omega$. Thus a dimension-4 parameter must be introduced to regularize this integral:

\begin{align}
    \Lambda^4 \equiv \int d\vec{k}
\end{align}

\noindent so that the effective action is

\begin{align}
    W \sim W_0 - \frac{\Lambda^4n^2}{2\lambda v^2}\int \frac{d^4x} {r^2}
\end{align}

\noindent plus higher order (integrable) corrections. The constant $W_0$ is just a volume divergence which may be ignored, as it is also present in the case of no string. Wick-rotating back, we find the effective Minkowski action:

\begin{align}
    iW = S \sim S_0 + \frac{\Lambda^4n^2}{2\lambda v^2}\int \frac{d^4x} {r^2}
\end{align}

\noindent Of course, for a static string solution the result is independent of time, and so the Hamiltonian receives a contribution from the vacuum effects of the form

\begin{align}
    H \supseteq -\frac{\Lambda^4n^2}{2\lambda v^2}\int dz \int d\theta \int \frac{dr}{r}
\end{align}

If one naively takes this expression and sets $\Lambda^4 =2\lambda v^4$, this divergent contribution to the Casimir energy exactly cancels the piece at classical order, rendering the string tension finite. In general however, the string tension is given by

\begin{align}
    T = 2\pi n^2v^2\int\frac{dr}{r}\left(1-\frac{\Lambda^4}{2\lambda v^4}\right)
\end{align}

\noindent plus finite/integrable terms higher order in $1/r$.

\section{An Effective Action Formulation}

There is a secondary approach to deriving the result of equation (24). Consider again the definitions of $W$ and $\Delta$: in particular, that $\ln\Delta$ can be written as the sum of two logarithms:

\begin{align}
   \ln\Delta &= \ln\left(-\partial_\mu(f^2\partial^\mu)\right)\\
   &= \ln(f^2) + \ln(-\partial^2-\partial_\mu\ln(f^2)\partial^\mu)
\end{align}

\noindent The operator in the second logarithm may be written as $-D^2+E$ for covariant derivative $D$ and connection $\omega$:

\begin{align}
    D_\mu = \partial_\mu + \omega_\mu\\
    \omega_\mu = \partial_\mu \ln(f)
\end{align}

\noindent and a potential 

\begin{align}
    E = \partial^2 \ln(f) + \left(\partial\ln(f)\right)^2
\end{align}

\noindent The primary method for analysis of the spectrum of this secondary operator should be the heat kernel approach, which yields an effective action which is a functional of $E$ and its covariant derivatives. The reader may check that the contributions to the effective action and therefore to the total energy from this secondary operator enter at $\mathcal{O}(r^{-4})$. The more important term is the first contribution. It yields an effective action of the form

\begin{align}
S\sim S_0-\frac{\Lambda^4}{2}\int d^4x \ln(f^2)
\end{align}

\noindent which in the large $r$ limit may be expanded, i.e.

\begin{align}
    \ln(f^2) \sim 1-\frac{n^2}{\lambda v^2 r^2}
\end{align}

\noindent yielding

\begin{align}
    S \sim S_0 + \frac{\Lambda^4n^2}{2\lambda v^2}\int \frac{d^4x} {r^2} + \text{constants}
\end{align}

\noindent identical to the previously derived result (22). The way $\Lambda^4$ enters into equation (30) is through the trace, which denotes a spacetime integration as well as a summation of all momentum modes, or in other words, equation (20).

\section{Discussion}

Clearly, $\Lambda$ is a UV cutoff on the momentum space integrals involving decoupled Goldstone Bosons. A priori, there is no justification for thinking it should have a particular value, however the result we have found is intriguing. We were partially motivated to preform this calculation by a recent lattice result\cite{2}. There, the authors investigated the global vortex, which is the $2+1$ dimensional version of the global string. The global vortex is a cross-section of the global string, and so classically, its mass diverges in the same was as the string tension. However, the lattice observation contradicted this: their strong coupling analysis found a finite mass for the vortex. Here we have demonstrated, as our calculation is also valid in the case of the model one dimension lower, that consideration of the infrared effects of the massless sector of the model can play a role in modifying the coefficient of that classically divergent mass/tension.\\

This suggests that a full analysis (perhaps involving resummation) of the infrared degrees of freedom, making use of the renormalization group, could yield an effective action for the string or vortex in which the infrared divergent contribution to the tension has been fully resolved and regulated away. We make no such guess as to what the form of this effective action might be, only conjecture that the issue of diverging tension will disappear.\\

The way in which the present analysis needs to be imporved is that it does not take into account that in a full effective theory of the string background: the string profile is likely to change and be deformed by this back-reaction from the Goldstone bosons. We did not apply the renormalization group to the analysis and therefore the exact result is not meant to be taken completely at face value. It should not be dismissed out of hand because of this, however. The existence of an infrared divergent piece of exactly the same form as the classical contribution to the tension is a surprise, and hints at the possibility of an effective theory in which the string tension can be made finite. Any such effective theory would be of great use to those who simulate string dynamics in the early Universe. While there is no particular reason to think $\Lambda^4 = 2\lambda v^4$ in the present analysis, there is no particular reason it cannot be larger than this value - it is an arbitrary momentum space cutoff - especially in the case of strong coupling ($\lambda >> 1$), where this cutoff is small compared to the symmetry breaking mass scale. If $\Lambda^4 >2\lambda v^4$, the tension appears to be infinite and negative though, signifying that the action of the string is certainly no longer a minima in field space, or even a saddle point. In this case, we should likely conclude this to be a sign of instability, or more exactly, evidence of a symmetry restoration phase at higher energies\cite{3,4,5}. \\

Not only does this work suggest the solitary global string ought to have finite tension, it also suggests that separated, oppositely wound strings need not be confined. As previously mentioned, the potential between a string of winding number $n$ and one of winding number $-n$ scales like the logarithm of their spatial separation. The analogous result to ours in the multi-string case suggests that this may not be the case after quantization. Of particular importance is the implications this has for cosmology: global strings are critical to the discussion of the axion\cite{6,7,8,9,10}, as they can form from the spontaneous breakdown of the $U(1)$ Peccei-Quinn symmetry\cite{11,12,13}. A signifigant portion of the current work involving axion dark matter involves the study of how these strings interact and ultimately decay into the currently observed dark matter densities. If the strings' long range interactions differ significantly from the classical field theory in the way we have suggested, it means that the number of strings formed per unit volume during the Peccei-Quinn phase transition could be much larger than previously expected. Not only that, but the resulting string dynamics and estimation of mean lifetime and relic axion abundance is also likely modified in a significant way. Besides the potential for the semi-classical analysis involving these strings to have mischaracterized the axion string tension and potential, we do not have anything meaningful to say about this. We have however provided evidence that a far more detailed analysis of the infrared degrees of freedom for the global string is necessary, and that this analysis has the potential to reveal a much different theory than what is worked with at the classical or semi-classical level. And as pointed out previously by \cite{2} in the case of the global vortex, this result potentially constitutes an invalidation of Derrick's theorem in the quantum theory.

\nocite{*}

\bibliography{ref}

\providecommand{\noopsort}[1]{}\providecommand{\singleletter}[1]{#1}%
\begin{thebibliography}{13}%
\makeatletter
\providecommand \@ifxundefined [1]{%
 \@ifx{#1\undefined}
}%
\providecommand \@ifnum [1]{%
 \ifnum #1\expandafter \@firstoftwo
 \else \expandafter \@secondoftwo
 \fi
}%
\providecommand \@ifx [1]{%
 \ifx #1\expandafter \@firstoftwo
 \else \expandafter \@secondoftwo
 \fi
}%
\providecommand \natexlab [1]{#1}%
\providecommand \enquote  [1]{``#1''}%
\providecommand \bibnamefont  [1]{#1}%
\providecommand \bibfnamefont [1]{#1}%
\providecommand \citenamefont [1]{#1}%
\providecommand \href@noop [0]{\@secondoftwo}%
\providecommand \href [0]{\begingroup \@sanitize@url \@href}%
\providecommand \@href[1]{\@@startlink{#1}\@@href}%
\providecommand \@@href[1]{\endgroup#1\@@endlink}%
\providecommand \@sanitize@url [0]{\catcode `\\12\catcode `\$12\catcode
  `\&12\catcode `\#12\catcode `\^12\catcode `\_12\catcode `\%12\relax}%
\providecommand \@@startlink[1]{}%
\providecommand \@@endlink[0]{}%
\providecommand \url  [0]{\begingroup\@sanitize@url \@url }%
\providecommand \@url [1]{\endgroup\@href {#1}{\urlprefix }}%
\providecommand \urlprefix  [0]{URL }%
\providecommand \Eprint [0]{\href }%
\providecommand \doibase [0]{https://doi.org/}%
\providecommand \selectlanguage [0]{\@gobble}%
\providecommand \bibinfo  [0]{\@secondoftwo}%
\providecommand \bibfield  [0]{\@secondoftwo}%
\providecommand \translation [1]{[#1]}%
\providecommand \BibitemOpen [0]{}%
\providecommand \bibitemStop [0]{}%
\providecommand \bibitemNoStop [0]{.\EOS\space}%
\providecommand \EOS [0]{\spacefactor3000\relax}%
\providecommand \BibitemShut  [1]{\csname bibitem#1\endcsname}%
\let\auto@bib@innerbib\@empty
\bibitem [{\citenamefont {Rajaraman}(1982)}]{1}%
  \BibitemOpen
  \bibfield  {author} {\bibinfo {author} {\bibfnamefont {R.}~\bibnamefont
  {Rajaraman}},\ }\href@noop {} {\emph {\bibinfo {title} {Solitons and
  Instantons}}}\ (\bibinfo  {publisher} {North-Holland},\ \bibinfo {year}
  {1982})\BibitemShut {NoStop}%
\bibitem [{\citenamefont {Delfino}\ \emph {et~al.}(2019)\citenamefont
  {Delfino}, \citenamefont {Selke},\ and\ \citenamefont {Squarcini}}]{2}%
  \BibitemOpen
  \bibfield  {author} {\bibinfo {author} {\bibfnamefont {G.}~\bibnamefont
  {Delfino}}, \bibinfo {author} {\bibfnamefont {W.}~\bibnamefont {Selke}},\
  and\ \bibinfo {author} {\bibfnamefont {A.}~\bibnamefont {Squarcini}},\
  }\href@noop {} {\bibfield  {journal} {\bibinfo  {journal} {Phys.\ Rev.}\
  }\textbf {\bibinfo {volume} {122}} (\bibinfo {year} {2019})}\BibitemShut
  {NoStop}%
\bibitem [{\citenamefont {Boniniac}\ and\ \citenamefont
  {D'Attanasiobc}(1996)}]{3}%
  \BibitemOpen
  \bibfield  {author} {\bibinfo {author} {\bibfnamefont {M.}~\bibnamefont
  {Boniniac}}\ and\ \bibinfo {author} {\bibfnamefont {M.}~\bibnamefont
  {D'Attanasiobc}},\ }\href@noop {} {\bibfield  {journal} {\bibinfo  {journal}
  {Nucl.Phys. B}\ }\textbf {\bibinfo {volume} {466}},\ \bibinfo {pages} {315}
  (\bibinfo {year} {1996})}\BibitemShut {NoStop}%
\bibitem [{\citenamefont {Coleman}\ and\ \citenamefont {Weinberg}(1973)}]{4}%
  \BibitemOpen
  \bibfield  {author} {\bibinfo {author} {\bibfnamefont {S.}~\bibnamefont
  {Coleman}}\ and\ \bibinfo {author} {\bibfnamefont {E.}~\bibnamefont
  {Weinberg}},\ }\href@noop {} {\bibfield  {journal} {\bibinfo  {journal}
  {Phys. Rev. D}\ }\textbf {\bibinfo {volume} {7}} (\bibinfo {year}
  {1973})}\BibitemShut {NoStop}%
\bibitem [{\citenamefont {Polchinski}(1984)}]{5}%
  \BibitemOpen
  \bibfield  {author} {\bibinfo {author} {\bibfnamefont {J.}~\bibnamefont
  {Polchinski}},\ }\href@noop {} {\bibfield  {journal} {\bibinfo  {journal}
  {Nucl. Phys. B}\ }\textbf {\bibinfo {volume} {231}},\ \bibinfo {pages} {269}
  (\bibinfo {year} {1984})}\BibitemShut {NoStop}%
\bibitem [{\citenamefont {Peccei}\ and\ \citenamefont {Quinn}(1977)}]{6}%
  \BibitemOpen
  \bibfield  {author} {\bibinfo {author} {\bibfnamefont {R.~D.}\ \bibnamefont
  {Peccei}}\ and\ \bibinfo {author} {\bibfnamefont {H.~R.}\ \bibnamefont
  {Quinn}},\ }\href@noop {} {\bibfield  {journal} {\bibinfo  {journal} {Phys.
  Rev. Lett.}\ }\textbf {\bibinfo {volume} {38}} (\bibinfo {year}
  {1977})}\BibitemShut {NoStop}%
\bibitem [{\citenamefont {Kim}(1979)}]{7}%
  \BibitemOpen
  \bibfield  {author} {\bibinfo {author} {\bibfnamefont {J.~E.}\ \bibnamefont
  {Kim}},\ }\href@noop {} {\bibfield  {journal} {\bibinfo  {journal} {Phys.
  Rev. Lett.}\ }\textbf {\bibinfo {volume} {43}} (\bibinfo {year}
  {1979})}\BibitemShut {NoStop}%
\bibitem [{\citenamefont {Shifman}\ \emph {et~al.}(1980)\citenamefont
  {Shifman}, \citenamefont {Vainstein},\ and\ \citenamefont {Zakharov}}]{8}%
  \BibitemOpen
  \bibfield  {author} {\bibinfo {author} {\bibfnamefont {M.~A.}\ \bibnamefont
  {Shifman}}, \bibinfo {author} {\bibfnamefont {A.~I.}\ \bibnamefont
  {Vainstein}},\ and\ \bibinfo {author} {\bibfnamefont {V.~I.}\ \bibnamefont
  {Zakharov}},\ }\href@noop {} {\bibfield  {journal} {\bibinfo  {journal}
  {Nucl. Phys B.}\ }\textbf {\bibinfo {volume} {166}} (\bibinfo {year}
  {1980})}\BibitemShut {NoStop}%
\bibitem [{\citenamefont {Dine}\ \emph {et~al.}(1981)\citenamefont {Dine},
  \citenamefont {Fischler},\ and\ \citenamefont {Srednicki}}]{9}%
  \BibitemOpen
  \bibfield  {author} {\bibinfo {author} {\bibfnamefont {M.}~\bibnamefont
  {Dine}}, \bibinfo {author} {\bibfnamefont {W.}~\bibnamefont {Fischler}},\
  and\ \bibinfo {author} {\bibfnamefont {M.}~\bibnamefont {Srednicki}},\
  }\href@noop {} {\bibfield  {journal} {\bibinfo  {journal} {Phys. Lett. B}\
  }\textbf {\bibinfo {volume} {104}},\ \bibinfo {pages} {199} (\bibinfo {year}
  {1981})}\BibitemShut {NoStop}%
\bibitem [{\citenamefont {Zhitnitsky}(1980)}]{10}%
  \BibitemOpen
  \bibfield  {author} {\bibinfo {author} {\bibfnamefont {A.~R.}\ \bibnamefont
  {Zhitnitsky}},\ }\href@noop {} {\bibfield  {journal} {\bibinfo  {journal}
  {Sov. J. Nucl. Phys.}\ }\textbf {\bibinfo {volume} {31}} (\bibinfo {year}
  {1980})}\BibitemShut {NoStop}%
\bibitem [{\citenamefont {Kibble}(1976)}]{11}%
  \BibitemOpen
  \bibfield  {author} {\bibinfo {author} {\bibfnamefont {T.~W.}\ \bibnamefont
  {Kibble}},\ }\href@noop {} {\bibfield  {journal} {\bibinfo  {journal} {J.
  Phys. A}\ }\textbf {\bibinfo {volume} {9}} (\bibinfo {year}
  {1976})}\BibitemShut {NoStop}%
\bibitem [{\citenamefont {Kibble}(1980)}]{12}%
  \BibitemOpen
  \bibfield  {author} {\bibinfo {author} {\bibfnamefont {T.~W.}\ \bibnamefont
  {Kibble}},\ }\href@noop {} {\bibfield  {journal} {\bibinfo  {journal} {Phys.
  Rept.}\ }\textbf {\bibinfo {volume} {67}},\ \bibinfo {pages} {183} (\bibinfo
  {year} {1980})}\BibitemShut {NoStop}%
\bibitem [{\citenamefont {Vilenkin}\ and\ \citenamefont {Everett}(1982)}]{13}%
  \BibitemOpen
  \bibfield  {author} {\bibinfo {author} {\bibfnamefont {A.}~\bibnamefont
  {Vilenkin}}\ and\ \bibinfo {author} {\bibfnamefont {A.~E.}\ \bibnamefont
  {Everett}},\ }\href@noop {} {\bibfield  {journal} {\bibinfo  {journal} {Phys.
  Rev. Lett.}\ }\textbf {\bibinfo {volume} {48}} (\bibinfo {year}
  {1982})}\BibitemShut {NoStop}%
\end{thebibliography}%

\end{document}